\documentclass[graybox]{svmult}

\usepackage{helvet}         
\usepackage{makeidx}         
\usepackage{graphicx}        
\usepackage{multicol}        
\usepackage[bottom]{footmisc}
\usepackage{natbib}
\setcitestyle{aysep={}}
\usepackage{amssymb}
\usepackage{titlesec}
\usepackage{lipsum}

\def\dlambda{$\lambda\lambda$}
\def\kms{km~s$^{-1}$}

\begin{document}

%
%
%

\title*{The Supernova --- Supernova Remnant Connection}
\author{Dan Milisavljevic and Robert A.\ Fesen}
\institute{Dan Milisavljevic \at Harvard-Smithsonian Center for
  Astrophysics, 60 Garden St., Cambridge, MA 02138, \email{dmilisav@cfa.harvard.edu}
\and Robert A.\ Fesen \at Dartmouth College, 6127 Wilder Laboratory,
Hanover, NH 03755, \email{robert.fesen@dartmouth.edu}}

\maketitle

\abstract{Many aspects of the progenitor systems, environments, and
  explosion dynamics of the various subtypes of supernovae are
  difficult to investigate at extragalactic distances where they are
  observed as unresolved sources. Alternatively, young supernova
  remnants in our own galaxy and in the Large and Small Magellanic
  Clouds offer opportunities to resolve, measure, and track expanding
  stellar ejecta in fine detail, but the handful that are known
  exhibit widely different properties that reflect the diversity of
  their parent explosions and local circumstellar and interstellar
  environments. A way of complementing both supernova and supernova
  remnant research is to establish strong empirical links between the
  two separate stages of stellar explosions. Here we briefly review
  recent progress in the development of supernova---supernova remnant
  connections, paying special attention to connections made through
  the study of ``middle-aged'' ($\sim 10-100$ yr) supernovae and young
  ($< 1000$ yr) supernova remnants. We highlight how this approach can
  uniquely inform several key areas of supernova research, including
  the origins of explosive mixing, high-velocity jets, and the
  formation of dust in the ejecta.}

\section{Introduction}
\label{sec:intro}

Distant extragalactic supernovae (SNe) appear as unresolved point
sources. This inescapable fact severely restricts our ability to
investigate SNe in fine detail, and introduces a fundamental obstacle
to obtaining three-dimensional kinematic and chemical data about the
expanding ejecta that can be used to extract key properties of the
progenitor star system and the explosion processes associated with the
supernova.

One way around this problem is to examine young, nearby supernova
remnants (SNRs) that encode valuable information in their expanding
debris and evolution. Investigations of SNRs offer opportunities to
resolve, measure, and track the expanding stellar ejecta, which in
turn provides detailed information about the explosion dynamics,
nucleosynthetic yields, and mixing of the progenitor star's chemically
distinct layers. Additionally, the progenitor star system's
evolutionary stages and mass-loss can be explored by studying the SN's
interaction with its surrounding environment, and the fate of the SN's
remnant core can be deduced through observations of central compact
objects (CCOs).

However, SNR investigations come with their own set of
limitations. With increasing age, interpretations of debris fields are
complicated by the SN's increasing interaction with local
environments. Observed asymmetry in SN morphology may be due to an
inhomogeneous circumstellar and/or interstellar mediums (CSM/ISM), or
non-spherical explosion. Consequently, the number of suitable objects
of investigation where these types of effects are minimal is
relatively small.

Establishing links between the variety of extragalactic SNe seen and
the diverse collection of young, nearby and resolved SNRs is a way of
complementing both areas of research.  SN studies provide data on
diversity and rates of SN types and subclasses as well as peak mean
ejection velocities of chemically distinct layers, while SNRs provide
detailed three dimensional kinematic and chemical information of the
ejecta and progenitor star mass loss distribution.  Here we give a
brief overview of progress made at establishing connections between
SNe and SNRs.

\section{The Remnants of Historical Galactic Supernovae }
\label{sec:historical}
\index{historical supernovae}
\index{classification}

Although there are about 300 SNRs identified in the Milky Way, less
than a dozen are $\lesssim1000$ years old \citep{Green14}. Young
remnants are particularly useful for establishing SN--SNR connections
because they have not been diluted by interaction with their local
circumstellar and insterstellar mediums (CSM/ISM) and exhibit
properties still reflective of the type of SN from which they
originated. Especially young Galactic remnants have the added
advantage that many have been linked to historically witnessed
celestial events \citep{CS77}.

Modern classification of SNe follows a system based on the lack or
presence of spectral lines seen at optical wavelengths around the time
of explosion when the SN is most luminous. The primary division is
made between Type I and II events, first proposed by
\citet{Minkowski41}. Type I SNe lack conspicuous features associated
with hydrogen and Type II SNe show clear hydrogen lines. Decades of SN
discovery has uncovered a zoo of events of diverse spectral properties
and luminosities \citep{Fil97,Gal-Yam12}. The two major divisions of
SNe are now generally differentiated by the types of explosive
progenitor stars: thermonuclear explosions of white dwarfs (Type Ia)
and core collapses of massive ($> 8\;\rm M_{\odot}$) stars (Type II,
IIn, IIb, Ib, and Ic).

The far broader range of observed properties in core-collapse SNe
compared to SN Ia reflect both a wider range of progenitor masses and
the loss of a hydrogen and/or helium rich envelope at the time of
outburst \citep{Fil97}.  This diversity of high mass progenitors and
SN subtypes is in contrast to just three main types of young SNRs:
O-rich remnants like Cassiopeia A (Cas~A), pulsar dominated or
so-called plerion remnants like the Crab Nebula, and collisionless
shock dominated remnants like Tycho's SNR.

Early attempts at relating SN types with SN remnants involved
connecting SNRs with historical records of ``guest stars''
\citep{CS77}. Because these ancient sightings provide no information
about the spectroscopic properties of the events, and offer scant
and sometimes questionable information about the SN's peak
brightness, color, and duration of visibility
\citep{Schaefer96,Fesen12}, they are limited in their ability to
relate young Milky Way remnants to extragalactic SNe.

Most young galactic SNRs have strong individual characteristics that
are not always shared by other young SNRs or easily connected to SN
subclasses.  Such differences largely arise from the role CSM can play
in affecting a remnant's properties and evolution. For example, the
young galactic remnant Cas~A shows optical properties unlike most of
the other O-rich SNRs for which it is suppose to be the prototype. The
galactic remnant W50/SS433 and the O-rich ejecta + 50 msec pulsar
remnant 0540-69 in the LMC also defy the three simple cataloging
classes.

However, recent studies at X-ray wavelengths have provided reasonably
good methods for differentiating Type Ia remnants from remnants of
core-collapse SNe. These methods of analyses include estimates of iron
abundance \citep{Reynolds07}, X-ray line morphologies \citep{Lopez09},
and Fe-K line energy centroids \citep{Patnaude15}.  In some cases
these analyses are able to provide SN subclassifications
\citep{Patnaude12}. Unfortunately, in most cases Galactic remnants are
too old for such methods to provide this level of detail.

The famous Crab Nebula \index{Crab Nebula} is an example of the
difficulty of connecting a SNR with a particular type of extragalactic
SN.  Although from ancient records we know its precise age and peak
visual brightness, the historic SN of 1054 AD left a remnant unlike
any other in our galaxy.  With an expansion velocity ($<2000$ \kms)
far lower than commonly seen in SNe, an extremely luminous pulsar, and
ejecta showing only helium enrichment -- and even then only in a select
band of filaments -- debate continues as to which SN subtype best
matches its properties \citep{DF85,Smith2013,YC15}.

Robust connections between various SN types and young SNRs have
recently been achieved through the study of SN light echos\index{light
  echoes}.  Light from the SN scattered by interstellar dust can be
observed after a time delay resulting from a longer path length.
Spectroscopy of these light echoes has led to robust SN associations
between the young Milky Way and LMC/SMC remnants 0509-67.5 and Tycho
with SN Ia events \citep{Rest08,Krause08Tycho}.  Similarly, Cas~A has
been associated with the Type IIb SN\,1993J
\citep{Krause08CasA,Rest11}, which is discussed in detail in
Section~\ref{sec:connection}.

\section{The SN to SNR Transition}
\label{sec:transition}
\index{SN to SNR transition}

There is no generally accepted definition for the point when a SN
becomes a SNR.  A theoretical definition is that the remnant phase
begins when a SN departs from free expansion and begins to strongly
interact with its surrounding ISM. On the other hand, an operational
definition might be when both UV/optical line and continuum emission
from a SN's ejecta falls below that generated by interaction with
either surrounding CSM/ISM material or via emission from a central
compact stellar remnant \citep{Fesen01SNR}.  However, these
definitions are not applicable for Type IIn SNe that exhibit strong
SN-CSM interaction immediately after explosion and may represent
$\sim 10$\% of core-collapse explosions \citep{Smith11}.

While monitoring the evolution of a SN from explosion into a young
remnant over timescales of $10^2 - 10^3$ yr is not practical,
timescales of decades are short enough to permit SN--SNR connections.
This was first recognized in the late 1980s with the optical
re-detections of SN~1980K \citep{FB88} and SN~1957D \citep{Long89},
and on-going studies of SN 1987A in the LMC have been helpful in
allowing us to witness certain aspects of this identity
change. However, the particular properties of any one intermediate- or
``middle''-aged SN do not offer the desired broad insights about the
formation and evolution of young remnants across the whole spectrum of
SN subclasses. To accomplish this, examinations of a variety of
well-evolved SNe is required.

\begin{figure}[t]

\includegraphics[width=\linewidth]{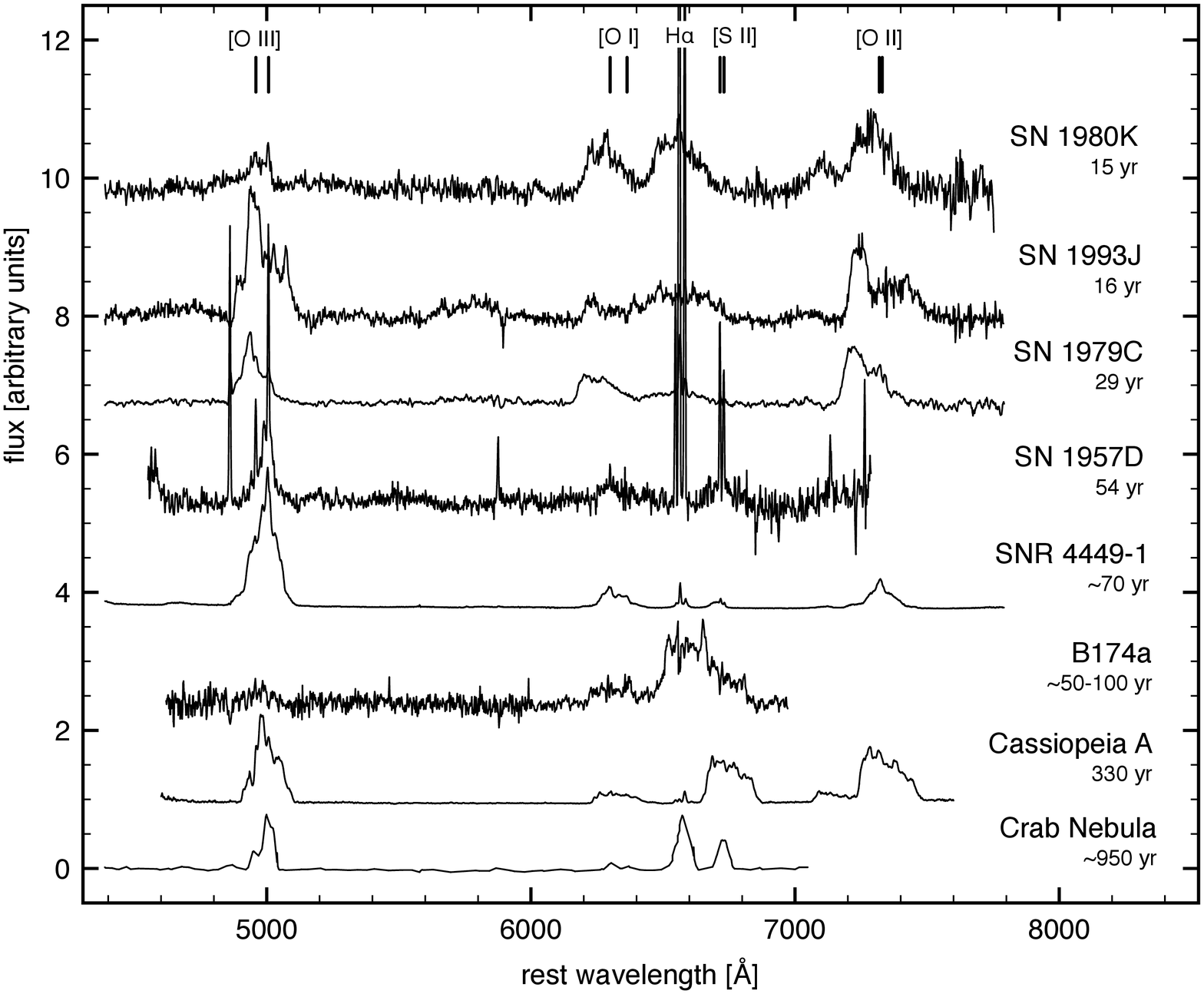}

\caption{Several well-observed middle-aged SNe and two spatially
  integrated spectra of young SNRs. Data have been originally
  published in the following papers: SN\,1979C
  \citep{Milisavljevic09}; SN\,1957D \citep{Long12}; B174a
  \citep{Blair15}; SNR\,4441-1 \citep{MF08}; Crab Nebula
  \citep{Smith03}; all others are from \citet{Milisavljevic12}.}

\label{fig:spectra}     
\end{figure}

Detections of SNe passing through the transitional state into remnants
are relatively rare. This is chiefly a consequence of their fairly
rapid decline in luminosity.  Because the majority of SNe occur at
distances $>10$ Mpc and fade at least eight magnitudes below peak
brightness in their first two years, observations have been largely
limited to one year or so after maximum light when they are at
apparent magnitudes $< 20$.

Fortunately, favorable circumstances sometimes make it possible to
monitor SNe many years or even decades post-outburst. In the special
case of SN~1987A, its close proximity in the LMC has enabled long-term
monitoring. In the majority of other cases when late-time monitoring
is possible, some long-lived energy source related to interaction
between the SN and ISM/CSM or a CCO maintains optical luminosity at
observable levels. Because young Type Ia remnants do not appear to undergo
strong ISM or CSM interactions and do not contain CCOs, the majority
of recent SN--SNR studies have concentrated on core-collapse
SNRs. Accordingly, this chapter focuses on core-collapse
events. Exceptional cases of middle-aged Type Ia observations do
exist, however, such as G$1.9+0.3$ \citep{Borkowski14}, and SN\,1885,
which we highlight in Section \ref{sec:sn1885}.

Figure~\ref{fig:spectra} shows a sample of well-observed late-time
emissions (i.e., ``nebular spectra''\index{nebular spectra}) of
core-collapse SNe. The figure has been modified from
\citet{Milisavljevic12} to include the recent discovery of a very
young SNR by \citet{Blair15} during a survey of M83, and a superior
deep optical spectrum of SN\,1957D by \citet{Long12}.  Also shown in
the figure are spatially integrated spectra of two young supernova
remnants: Cas~A \citep{Milisavljevic12} and the Crab Nebula
\citep{Smith03}.

As shown in Figure~\ref{fig:spectra}, there is great variety in the
late-time optical spectra of middle-aged core-collapse SNe, seen
mostly in the relative line strengths and expansion
velocities of [O~I] \dlambda 6300, 6364,
[O~II] \dlambda 7319, 7330, [O~III] \dlambda 4959, 5007, [S~II]
\dlambda 6716, 6731, and H$\alpha$. Nonetheless,
important trends that reflect fundamental properties of these systems
do emerage upon close inspection. We discuss these trends in the next
section.

\begin{figure}[tp!]
\centering
\includegraphics[width=0.8\linewidth]{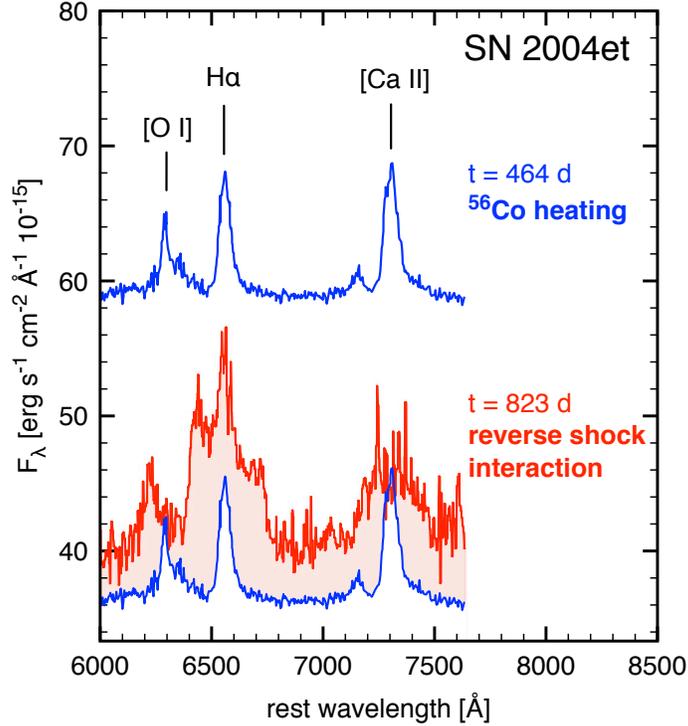}

\caption{Optical spectra of SN\,2004et from \citet{Kotak09}. At t =
  464 d, the emission from the [O~I], H$\alpha$ and [Ca~II] lines are
  powered by radioactive $^{56}$Co, which was originally $^{56}$Ni
  synthesized in the explosion and mixed from interior regions. At t =
  823 d, the emission line profiles of these ions undergo a dramatic
  broadening, from $V \sim 3000$ \kms\ to $V \sim 8500$ \kms, which is
  consistent with the reverse shock exciting outer, higher velocity
  ejecta. Emission around [Ca~II] at t = 823 d may have a contribution
  from [O~II].}

\label{fig:04et}      
\index{SN to SNR transition}
\end{figure}

\section{Energy Sources of Late-time Emission}
\label{sec:energy}
\index{nebular spectra}

Of the various late-time mechanisms theorized to power SN emission at
epochs $> 2$ yr, the most common process is the forward shock front
and SN ejecta interaction with surrounding CSM shed from the
progenitor star (see \citealt{CF03} and references therein).  These
SN--CSM interactions \index{SN--CSM interaction} lead to the formation
of a reverse shock \index{reverse shock} moving back into the
expanding ejecta, which may subsequently ionize a broad inner ejecta
region from which the optical lines are produced. Other proposed
late-time mechanisms include pulsar/magnetar interaction with
expanding SN gas \citep{CF92,Woosley10} or accretion onto a black-hole
remnant \citep{Patnaude11}.

The point at which SN--CSM interaction leading to reverse
shock-heating of inner ejecta dominates observed optical emission
ranges widely. For most core-collapse SNe, the timescale is at least
$1-2$ yr. The timescale is dominated by the dynamics of the explosion
and the local CSM/ISM environment it is running into. Prior to the
initiation of SN--CSM interaction, emission is dominated by inner
metal-rich ejecta that is excited from radioactive decay of
$^{56}$Ni$\rightarrow$ $^{56}$Co $\rightarrow$ $^{56}$Fe.

The SN--SNR transition between radioactive $^{56}$Co heating of O-rich
ejecta from the interior (directed outward) to reverse shock
excitation from the exterior (directed inward) is illustrated in
Figure~\ref{fig:04et}. Two epochs of late-time spectra of SN~2004et
\index{SN 2004et} (originally presented in \citealt{Kotak09}) show a
substantial increase in the width of the emission line profiles of
[O~I], [Ca~II], and H$\alpha$. This transformation is best understood
as the consequence of the reverse shock beginning to strongly interact
with metal-rich ejecta. Such an abrupt change in emission is rarely
observed in SNe but is an important step in the transition from SN to
SNR.

\begin{figure}[t]
\centering
\includegraphics[width=0.75\linewidth]{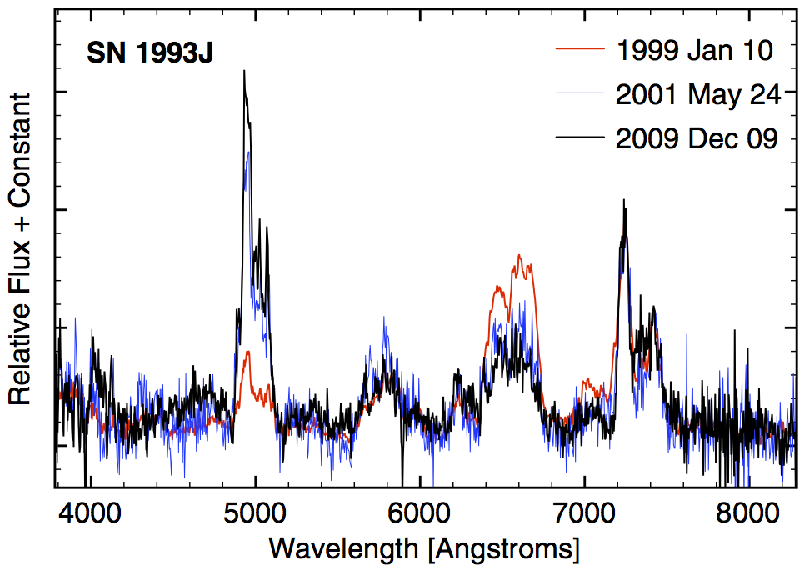}
\includegraphics[width=0.8\linewidth]{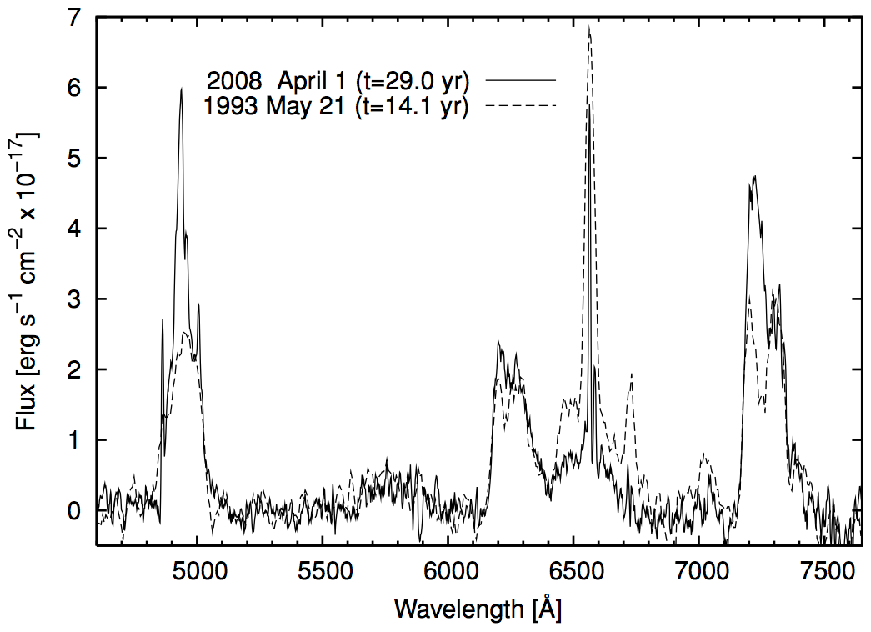}

\caption{Emission evolution in middle-aged SNe. Top: Three epochs of
  optical spectra of the Type IIb SN\,1993J \index{SN 1993J} showing
  the evolution of relative line strengths
  \citep{Milisavljevic12}. Bottom: Two epochs of optical spectra of
  the Type IIL SN\,1979C \index{SN 1979C}
  \citep{Milisavljevic09}. Declining H$\alpha$ emission with
  simultaneous increasing [O~III] emission is apparent, which is
  commonly observed in late-time emissions of SNe as ejecta continue
  to expand and cool and the reverse shock penetrates through the
  H-rich layers of the ejecta.}

\label{fig:evol}     
\end{figure}

The various mechanisms of late-time emission can be distinguished via
observed changes in velocity widths of emission line profiles
\citep{CF92,CF94}.  In circumstellar interaction scenarios where the
reverse shock penetrates into deeper layers of ejecta, the velocity
widths of emission lines are expected to narrow with time. Changes in
relative line strengths are also anticipated as expansion drives down
ejecta density. The flux ratio [O~III]/([O~I]+[O~II]) increases with
time and H$\alpha$/([O~I]+[O~II]) should decrease
(Figure~\ref{fig:evol}). Alternatively, in scenarios involving a
pulsar wind nebula where emission is powered by photoionization, line
widths are anticipated to broaden because of acceleration by the
pulsar wind bubble.

The extragalactic SNe shown in Figure~\ref{fig:spectra} all exhibit
the anticipated changes in emission line strengths and widths
attributable to SN--CSM interaction at optical wavelengths. However,
some of these SNe show signs of hosting pulsar wind nebulae or black
holes at X-ray wavelengths. For example, the X-ray flux density of
SN\,1970G \index{SN 1970G} recently increased by a factor of $\sim 3$
while its radio flux significantly lowered \citep{Dittmann14}. This
might be due to the turn-on of a pulsar wind nebula\index{pulsar wind
  nebula}. Similarly, SN\,1957D \index{SN 1957D} was recently
re-detected at X-ray wavelengths showing a hard and highly
self-absorbed spectrum which suggests the presence of an energetic
pulsar and its pulsar wind nebula \citep{Long12}. Perhaps most
remarkable is the constant X-ray luminosity of SN\,1979C \index{SN
  1979C} that possibly signals emission powered by either a
stellar-mass ($5-10\;\rm M_{\odot}$) black hole \index{black hole}
accreting material or a central pulsar wind nebula \citep{Patnaude11}.

\begin{figure}[t]
\centering
\includegraphics[width=0.6\linewidth]{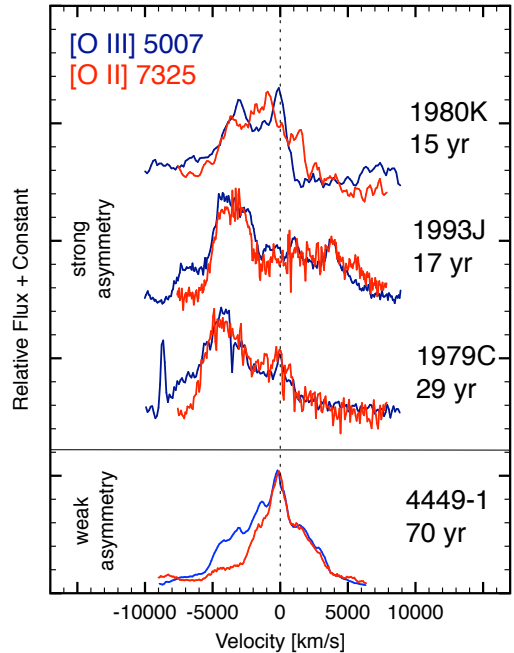}

\caption{Strong and weak emission line asymmetry in middle-aged
  SNe. Velocities of [O~III] 4959, 5007 emission line profiles are
  with respect to 5007~\AA. Velocities of [O~II] 7319, 7330 emission
  line profiles are with respect to 7325~\AA. Adapted from
  \citet{Milisavljevic12}.}

\label{fig:asymmetry}     
\end{figure}

\section{Emission Asymmetry}
\label{sec:asymmetry}

Asymmetric emission line profiles are common in late-time
core-collapse SN spectra. In Figure~\ref{fig:asymmetry}, we show three
examples of emission asymmetry in the [O~III] \dlambda 4959, 5007 line
profiles of SN\, 1980K, SN\,1993J, and SN\,1979C. The blue-red
asymmetry (i.e., most emission is observed at blueshifted velocities)
tends to be strongest in the forbidden oxygen lines, but can also be
observed in H$\alpha$, [S~II], and [Ar~III] line profiles. Hubble
Space Telescope ({\sl HST}) observations of SN\,1979C have shown that
this asymmetry can be even stronger at UV wavelengths
\citep{Fesen99}. A small subset of SNe do not exhibit pronounced
asymmetry in their emission line profiles, or exhibit a mixed
combination of profiles that both do and do not show asymmetry. For
example, the ultraluminous oxygen-rich SNR 4449-1 \index{SNR 4449-1}
shows mild asymmetry in its oxygen emission, but strong blue asymmetry
in its [S~II] \dlambda 6716, 6731 and [Ar~III] $\lambda$7136 emission
profiles with blueshifted velocity distributions spanning
$-2500 < V_{\rm exp} < 500$ \kms\ \citep{MF08,Bietenholz10}.

These asymmetric oxygen emission line profiles indicate that internal
absorption is obscuring emission from the SN's receding rear
hemisphere. Dust \index{dust} formation in the metal-rich ejecta is a
principal cause of this type of absorption at late
epochs. Alternatively, absorption by dust may take place in the cool
dense shell (CDS) in the shock region \citep{Deneault03}.  This is a
likely place for dust formation because of its high density and low
temperature \citep{CF94}. These emission line asymmetries can be
modeled to estimate dust composition and mass. For example,
\citet{BB16} modeled the late-time H$\alpha$ and [O~I] line emission
profiles of SN 1987A. Their results confirmed a steady increase in the
dust mass in SN 1987A's ejecta, from
$< 3 \times 10^{-3} \rm\; M_{\odot}$ on day 714 to
$> 0.1 \rm\; M_{\odot}$ by day 3604 (but see also
\citealt{DA15}). This time-dependent dust mass growth is relevant to
understanding the extent to which core-collapse SNe contribute to the
high quantities of dust observed in some high redshift galaxies
\citep{Watson15}.

In addition to the blue-red emission asymmetry, oxygen line profiles
of extragalactic SNe also show conspicuous emission line
substructure. This substructure, which comprises minor emission peaks
that can last many years ($> 10$), is typically interpreted as
``blobs'' or ``clumps'' \index{clumping} of material. Their presence
and longevity suggests that they are associated with regions of
emitting material that represent a large fraction of the entire
optically-emitting SN.

Large clumps are consistent with aspheric SN explosions with turbulent
mixing. Theory and observation strongly favor the notion that
asymmetric explosions drive core-collapse SNe.  However, where and how
this asymmetry is introduced is uncertain. Some explosion asymmetries
are potentially introduced by dynamical instabilities and the
influences of rotation and magnetic fields
\citep{Akiyama03,Janka12,Burrows13}, while others may be introduced by
a perturbed progenitor star structure \citep{AM11,CO13,Wong15}. In the
next section we explore how the robust SN--SNR connection between
SN\,1993J and Cas~A has provided powerful insight into the physical
origins of emission line asymmetry and substructure.

\section{The Cassiopeia A---SN\,1993J Connection}
\label{sec:connection}
\index{Cassiopeia A}
\index{SN 1993J}

Cas~A is considered the prototype for the class of young, oxygen-rich SNRs and
provides a clear look at the explosion dynamics of a core-collapse
SN. Cas A's distance of 3.4 kpc \citep{Reed95} has permitted detailed
study of its composition and distribution of SN ejecta on
fine scales, and its estimated current age of $\approx$340 yr
\citep{TFV01,Fesen06} places it at a stage of evolution not that
different from middle-aged SNe (Fig.\ \ref{fig:spectra}).

Cas A probably originated from a red supergiant progenitor with mass
$10-30$ M$_{\odot}$ that may have lost much of its hydrogen envelope
to a binary interaction \citep{CO03,Young06}.  Spectroscopy of optical
echoes of the SN outburst indicate that the original SN of Cas A was
of Type IIb \citep{Krause08CasA} and showed strong evidence for
asymmetry in the explosion \citep{Rest11}.  The integrated spectrum of
the Cas~A SNR is similar to many of the middle-aged CCSNe exhibiting
strong [O~III] emission. Particularly well-matched with Cas~A are
SN~1979C, SNR 4449-1, SN~1980K, and SN\,1993J.

\begin{figure}[t]

\includegraphics[width=\linewidth]{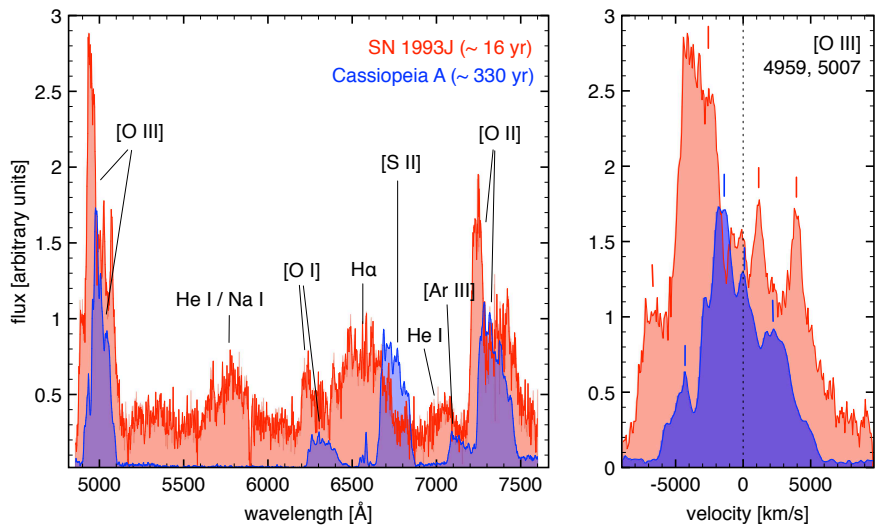}

\caption{Spatially integrated optical spectrum of Cas~A compared to
  that of SN\,1993J (modified from \citealt{Milisavljevic12}). The
  left panel shows the entire spectrum with dominant emission features
  identified. The right panel shows an enlargement around the [O~III]
  emission line profile. Solid lines highlight long-lasting emission
  line substructure.}

\label{fig:casa-93J}     
\end{figure}

In Figure~\ref{fig:casa-93J}, the optical spectra of Cas~A and
SN\,1993J are compared directly. Also shown in
Figure~\ref{fig:casa-93J} are the [O~III] emission line profiles
enlarged. Conspicuous minor emission peak substructure common to the
two spectra is highlighted. The differences and similarities between
the two are of especial interest given that Cas~A's Type IIb
classification was made by comparing its light echo spectra to that of
SN\,1993J near maximum light \citep{Krause08CasA,Rest11}. Together,
the two spectra illustrate the evolution of a Type IIb explosion from
decades to centuries after explosion. Three key properties of the
evolution reflect the reverse shock's progression toward the inner
layers of the ejecta: 1) The strength of H$\alpha$ emission decreases
as the H-rich outer layers of the ejecta are overtaken. 2) The
velocity of the emission line profile decreases. 3) Sulfur emission is
visible only after the reverse shock has penetrated into the inner
S-rich layers of the debris.

Both Cas~A and SN\,1993J show excess emission at blueshifted
velocities. Resolved multi-wavelength observations of Cas~A suggest that
this emission asymmetry is the result of absorption due to dust in the
ejecta. In particular, infrared studies of Cas~A have estimated that
upwards of $0.05\;\rm M_{\odot}$ of dust reside in its shocked ejecta
\citep{Rho08}. Thus, the same mechanism likely also applies for
SN\,1993J. This SN--SNR connection is consistent with notion that that
dust formation processes are significant when the reverse shock starts
to excite ejecta in the first few years post-explosion, and last for
centuries later.

\begin{figure}[t]

\includegraphics[width=\linewidth]{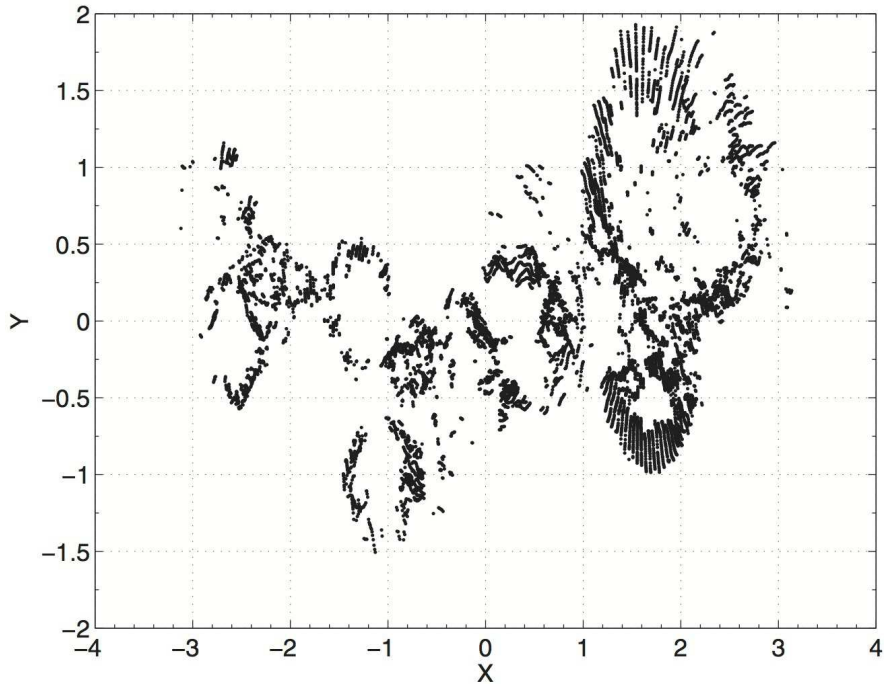}

\caption{The main shell of Cas A's reverse-shocked optically emitting
  ejecta as represented in a Mercator projection. At least six rings
  are observed. The largest ring located in the northwest is
  approximately twice the size of the other rings. The cylindrical map
  projection distorts the size and shape of large objects, especially
  toward the poles. Adapted from \citet{MF13}.}

\label{fig:mercator}      
\end{figure}

The young age of Cas~A and minimal influence of SN--CSM interaction has
kept its ejecta in ballistic trajectories from a well established
center of expansion \citep{TFV01}. This unique property has enabled
three dimensional reconstructions of the remnant at optical
\citep{Reed95,MF13}, and infrared and X-ray wavelengths
\citep{DeLaney10}. \citet{MF13} conducted a high-resolution optical
survey of Cas~A that included its highest velocity ejecta in the NE
and SW outflows sometimes called ``jets'' (see Section
\ref{sec:jets}). They found that the majority of the optical ejecta
are arranged in several well-defined and nearly circular ring-like
structures with diameters between approximately 30$''$ (0.5 pc) and
$2'$ (2 pc) (see Figure~\ref{fig:mercator}).

In light of the many connections between Cas~A and SN\,1993J, the
emission line substructure shared between them likely results from a
common origin. If so, then the ``clumps'' \index{clumping} of material
often observed in late-time spectra of core-collapse SNe are not
random, but instead associated with multi-ringed distributions of
ejecta like that seen and resolved in Cas A.  Strong evidence in support of this shared
morphology comes from kinematic reconstructions of other young SNRs
such as 1E 0102.2-7219 \index{1E 0102.2-7219} \citep{Eriksen01} and
N132D \index{N132D} \citep{Vogt11}, which also exhibit ejecta arranged
in ring-like geometries.

\section{Ejecta Bubbles and Large-Scale Mixing }
\label{sec:bubbles}
\index{Ni-bubbles}

\begin{figure}[t]
\centering
\includegraphics[width=0.9\linewidth]{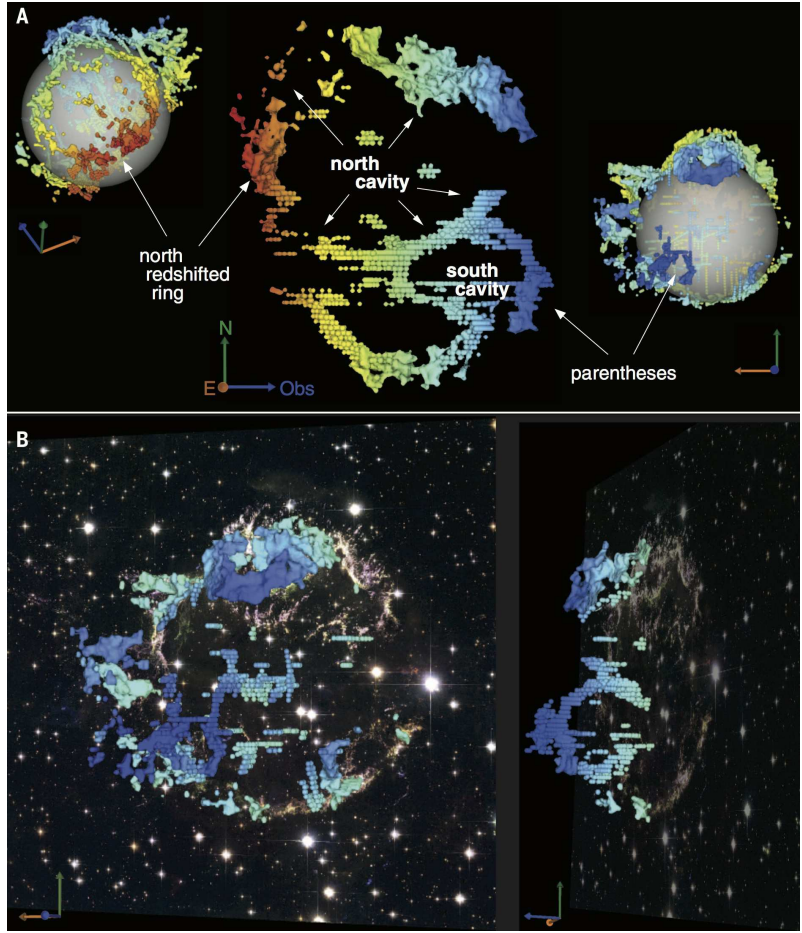}

\caption{Doppler reconstruction of Cas A. The blue-to-red color
  gradient corresponds to Doppler velocities that range from $-4000$
  to $+6000$ \kms. (A) A side perspective of a sliced portion of the
  remnant chosen to emphasize two conspicuous interior cavities and
  their connections to main-shell ejecta. The translucent sphere
  centered on the origin of expansion is a visual aid to differentiate
  between front and back material. (B) Two angled perspectives
  highlighting the south cavity. Adapted from \citet{MF15}. }

\label{fig:bubbles}     
\end{figure}

A near-infrared spectroscopic survey of Cassiopeia A's interior
unshocked S-rich material recently uncovered a bubble-like morphology
that smoothly connects with the multi-ringed structures seen in the
bright reverse shocked main shell of expanding debris
(Figure~\ref{fig:bubbles}; \citealt{MF15}).  These bubbles were most
likely created early in the explosion from plumes of radioactive
$^{56}$Ni-rich ejecta that decayed into cobalt and iron, releasing
energy that would cause the gas to expand and compress nearby
non-radioactive material such as oxygen, sulfur, and argon into cavity
walls. Doppler mapping of the X-ray bright Fe-K emission in Cas~A has
shown that three regions of Fe-rich material sits within three of
these rings \citep{DeLaney10,MF13}, which is consistent with this
proposed origin of the bubbles.

Compelling evidence for large-scale mixing \index{mixing} involving
considerable non-radial flow and the development of ``Ni-bubbles'' was
first observed in SN\,1987A. In that case, high energy gamma-rays and
X-rays with broad emission line widths from the decay of $^{56}$Ni
were detected only months after the explosion, implying that Ni-rich
material was near the star's surface well before 1D progenitor models
had predicted assuming spherical symmetry \citep{PW88,LMS93}.

Since SN 1987A, state-of-the-art 3D computer simulations of
core-collapse explosions have confirmed that large-scale mixing
originating from uneven neutrino heating can lead to Ni-dominated
plumes overtaking the star's outer oxygen- and carbon-rich layers with
velocities up to 4000 \kms\ \citep{Hammer10}. However, the majority of
these simulations show that the mass density should essentially be
unaffected. Though mixing affects the species distribution, the bulk
of the Ni mass should remain inside the remnant with velocities below
2000 \kms. This is, in fact, opposite to what is seen in Cas A, where
the X-ray bright Fe has velocities around and above the 4000 \kms\
limit. Thus, either the simulations are not adequately following the
dynamics of mixing, or more Fe remains to be detected in Cas A's
interior. This latter scenario is somewhat problematic. Approximately
$0.1\;\rm M_{\odot}$ of Fe (the decay product of Ni) is presently
observed in Cas~A \citep{Hwang12}, and yet the total amount produced
in the original SN is believed to have been less than this
\citep{Eriksen09}.

An additional consideration in interpreting a SN debris field is the
chemical make-up of the star at the time of outburst. The evolution of
massive stars toward the ends of their life cycles is likely to be
non-spherical and may have extensive inter-shell mixing. If strong
enough, these dynamical interactions lead to Rayleigh-Taylor
instabilities in the progenitor structure that can influence the
overall progression of the explosion and contribute to the formation
of Ni-rich bubbles \citep{AM11}. Thus, asymmetries introduced by a
turbulent progenitor star interior in addition to those initiated by
the explosion mechanism may contribute to bubble-like morphologies.

\section{Jet Activity in Core-collapse Supernovae}
\label{sec:jets}
\index{jets}

Interestingly, a continuum of explosion energies extending from
broad-lined Type Ic SNe associated with gamma-ray
bursts,\index{gamma-ray bursts} to more ordinary Type Ib/c SNe appears
to exist \citep{Soderberg10}. Such events suggest that a wide variety
of jet activity may potentially be occurring at energies that are
hidden observationally \citep{Margutti14,Milisavljevic15}. In these
cases, the central engine activity stops and becomes ``choked'' before
the jet is able to pierce through the stellar envelope. SNe associated
with choked jets lack sizable amounts of relativistic ejecta and thus
can be dynamically indistinguishable from ordinary core-collapse SNe
\citep{Lazzati12}.

Whether or not an internal jet successfully emerges from a star
experiencing a SN explosion is dependent on many factors. Perhaps the
most important factor is core rotation at the time of core
collapse. For the majority of SNe, rotation effects are anticipated to
be small and the explosion neutrino-driven. However, in a small subset
of cases where core rotation of the pre-SN star is rapid, magnetic
fields will be amplified and make magnetohydrodynamic
\index{magnetohydrodynamics} (MHD) power influential
\citep{Akiyama03}. In these extreme cases where the rotation rate is
very fast, MHD processes may dominate and a hypernova and/or a GRB
could result \citep{Burrows07}. Another factor is the progenitor star
size and composition. Large stellar envelopes and/or He layers may
inhibit central jets from completely piercing the surface of the star
(see, e.g., \citealt{Mazzali08}).

Searches for vestiges of prior SN jet activity in SNRs have had
largely inconclusive and/or debated conclusions. For example, the Crab
\index{Crab Nebula} has a curious filamentary feature that is 45$''$
wide and extends approximately 100$''$ off the nebula's northern
limb. Whether this ``jet'' was strongly sculpted by the surrounding
ISM or more directly influenced by the central pulsar has not been
resolved. Most recently, \citet{BF15} examined the Crab's 3D
filamentary structure along the jet's base and found a large and
nearly emission-free opening in the remnant's thick outer ejecta
shell. The jet's blueshifted and redshifted sides are surprisingly
well defined and, like the jet's sharp western limb, appear radially
aligned with the remnant's center of expansion.

Another example is the SNR W49B\index{W49B}.  \citet{Lopez13} conclude
that the morphological, spectral, and environmental characteristics of
W49B are indicative of a bipolar Type Ib/Ic SN explosion. However, the
arguments in favor of a jet-induced explosion for this fairly evolved
remnant that is known to be interacting with a nearby molecular cloud
are largely circumstantial. Optical and near-infrared spectroscopy has
been unable to detect any emission attributable to high velocity
metal-rich ejecta that could provide direct confirmation of a
jet-driven explosion.

On the other hand, Cas~A exhibits exceptionally high velocity Si- and
S-rich material in a jet/counter-jet arrangement
\citep{Fesen01,Hwang04}. The known extent of this jet region contains
fragmented knots of debris traveling $\sim 15,000$ \kms, which is
three times the velocity of the bulk of the O- and S-rich main
shell. Although the large opening half-angle of this high-velocity
ejecta is inconsistent with a highly collimated flow
\citep{Fesen01,MF13}, and spatial mapping of radioactive $^{44}$Ti
emission appears to rule out a jet-induced bipolar explosion
\citep{Grefenstette14}, some jet-like mechanism carved a path allowing
interior material from the Si-S-Ar-Ca region near the core out past
the mantle and H- and He-rich photosphere \citep{FM16}.

Given that the energy associated with the NE/SW jets of Cas~A is below
the energy anticipated to be associated with the original SN
explosion, an observational signature of their presence would be
hidden at extragalactic distances. The same is true for the Crab
Nebula. Only through nearby, resolved inspection is the existence of
these structures recognizable.

\section{The SN\,1885---SN Ia Connection}
\label{sec:sn1885}
\index{SN 1885}
\index{S And}

In contrast with core-collapse SNe that produce CCOs and
undergo strong ISM/CSM interactions that can energize observable
levels of late-time emissions, normal Type Ia SNe completely
disrupt their progenitor white dwarf stars and explode in relatively
``clean'' environments. Thus, after only a few years of free
expansion, the ejecta cool adiabatically and become effectively
invisible via line emission at extragalactic distances. This largely
explains why late-time optical detections of Type Ia SNe are
rare.

However, in exceptional circumstances, it has been possible to observe
late-time {\it absorptions} from middle-aged extragalactic Type Ia
SNe. This possibility was first realized with the re-detection of SN
1885 by \citet{Fesen89}. SN\,1885 was a bright historical nova (also
known as S And) discovered in late August of 1885 and located at a
projected distance of 16$''$ away from the nucleus of M31. The optical
spectrum, colors, and light curve evolution have suggested SN\,1885 to
be a subluminous Type Ia SN \citep{deV85}, although this
classification is uncertain \citep{Perets11}. SN\,1885's advantageous
close proximity to Andromeda's central bulge stars made it possible to
observe the remnant's ejecta via resonance line absorption.

\begin{figure}[tp]
\centering
\includegraphics[width=\linewidth]{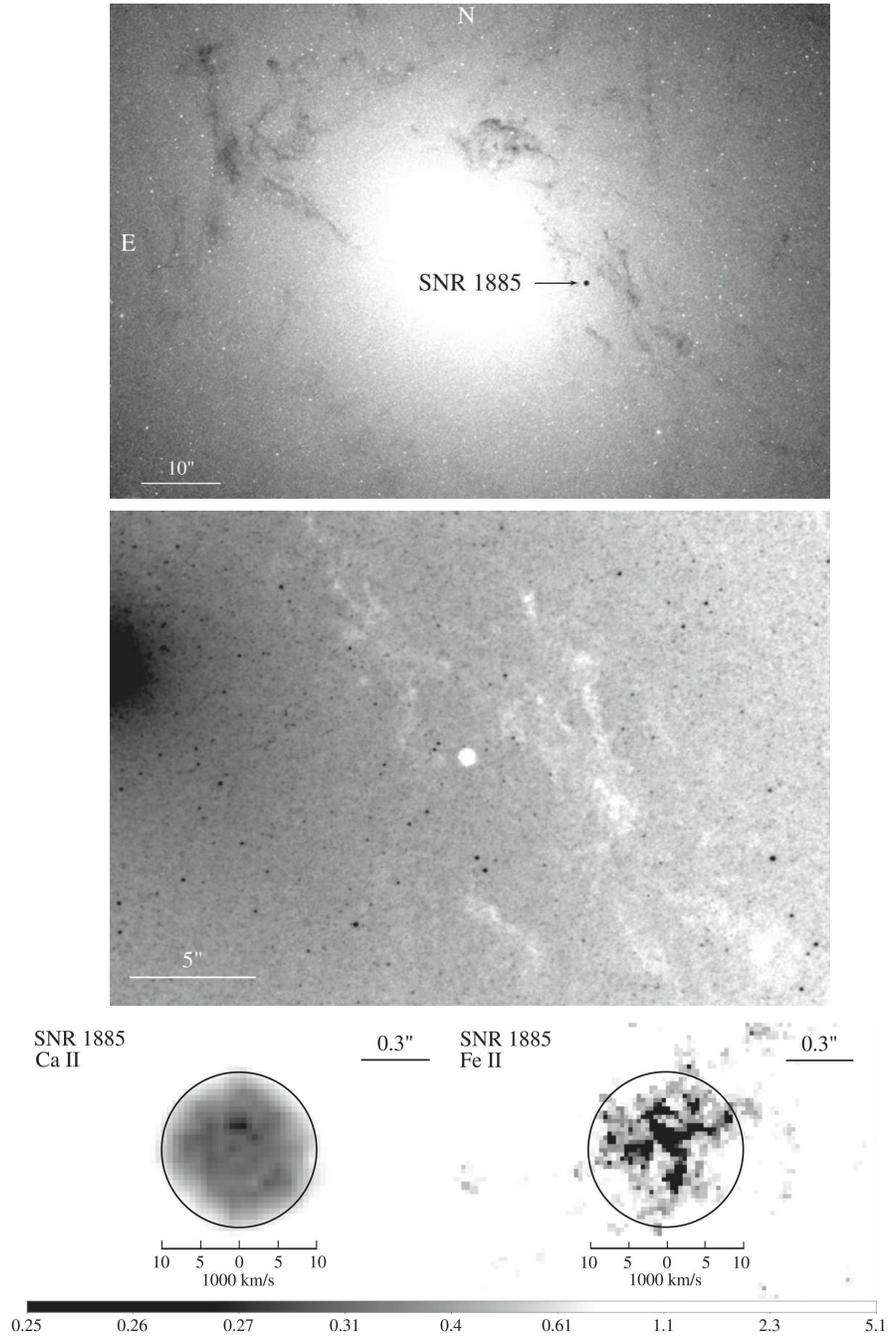}

\caption{{\sl HST} observations of SN\,1885. The top panel shows an
  WFC3 image of M31 taken with the F390M filter and displayed using a
  positive linear stretch. The location of SN\,1885, which appears as
  round dark spot of Ca~II H and K absorption, is marked. The middle
  panel shows an enlarged section of the same image centered on the SN
  and displayed using a negative log stretch. The bottom panels (left
  and right) show log intensity-stretched views of SN\,1885 as seen in
  Ca~II and Fe~II images. Adapted from \citet{Fesen15}. }

\label{fig:sn1885}     
\end{figure}

High resolution observations of SN\,1885 obtained with {\sl HST} have
been used to map the distribution of ejecta that are contained within
a $\approx 0.75''$ diameter (Figure~\ref{fig:sn1885}). Because the
debris are still in near-free expansion after 130 yr, they retain the
density distribution established shortly after the explosion and can
provide valuable kinematic information about the general properties
and character of a Type Ia explosion.

{\sl HST} Fe~II images of SN\,1885 uncovered four plumes of Fe-rich
material that extend from the remnant's center out to $\approx 10^4$
\kms. This distribution stands in strong contrast with the
distribution of Ca-rich material that is concentrated in a clumpy,
broken shell spanning velocities of $1000-5000$ \kms\ but extends out
to 12,500 \kms. \citet{Fesen15} find that the observed distributions
of ejecta are consistent with delayed detonation \index{delayed
  detonation} white dwarf models, and inconsistent with a highly
anisotropic explosion that could result from a violent merger
\index{double degenerate} of two white dwarfs.

\section{Conclusions}
\label{sec:conclusions}

In this chapter, we have reviewed how several important properties of
SN progenitor systems and explosion dynamics can be investigated
through connections between the many unresolved extragalactic SNe and
the few, resolved young SNRs seen in our galaxy, the Large and Small
Magellanic Clouds, and Andromeda.

Although observations of late-time optical emissions from Type Ia SNe
during the transitional middle-aged phase have not been possible, the
unique case of SN\,1885 has shown that the 2D arrangement of unshocked
Fe-rich and Ca-rich debris is most consistent with delayed detonation
white dwarf models. The plumes of Fe-rich material in this remnant may
explain the Fe II emission line asymmetries commonly seen in late-time
emission spectra of Type Ia SNe.

Middle-aged, core-collapse SNe powered by strong CSM interaction
leading to reverse shock-heating of ejecta are especially helpful in
making connections between SNe and SNRs.  A particularly important
SN--SNR connection is the Type IIb SN\,1993J in M81 and the 340 year
old O-rich galactic SNR Cassiopeia A.

Complex emission line substructures seen in SN\,1993J and other
core-collapse SNe typically interpreted as random ``clumps'' of
material may, in fact, be more easily understood as large-scale rings
of debris excited by a reverse shock like those seen in Cas A.  These
rings appear to be associated with a bubble-like interior originally
caused by outwardly expanding plumes of radioactive $^{56}$Ni-rich
ejecta.  Cas A's bipolar jet-like features is consistent with the
notion that a wide variety of jet activity may be occurring in
core-collapse SNe at energy scales that are observationally hidden at
extragalactic distances.

The SN--SNR connections made through decades-long monitoring of recent
SNe serve to emphasize the value of such observations.  While
high-quality late-time spectra are rare (less than 20 objects total),
analysis of these data can reveal important and sometimes unexpected
information. Furthermore, such SN--SNR connections can be used to
uniquely test our theoretical understandings of SNe. Only through
resolved inspections of SNRs can models of SNe be confronted with
empirical facts and gain insight into crucial processes that may be
lost in data obtained from unresolved extragalactic examples.

Finally, light echo spectroscopy has introducted a new and powerful
tool for developing robust SN--SNR connections.  It is anticipated
that increasing success obtaining optical spectra of light echoes
associated with historical Galactic SNe will continue to shed light on
remnants that don't quite fit our current SN classifications, such as
the Crab Nebula and Kepler. Morever, the growing awareness that
asymmetry plays a significant role in SN explosions provides strong
motivation for 3D light echo spectroscopy, which has the unique
capacity to yield information from multiple lines of sight.

\section*{Acknowledgments}

We thank B.\ Blair, K.\ Long, and F.\ Winkler for providing data of
B174a and SN\,1957D. A portion of this work was supported by NASA
through grants GO-12300 and GO-12674 from the Space Telescope Science
Institute (STScI), which is operated by the Association of
Universities for Research in Astronomy.

\section*{Cross-References}

\begin{itemize}

\item Historical records of supernovae

\item Supernova of 1054 and its remnant, the Crab Nebula

\item Supernova Remnant Cassiopeia A

\item Observational Classification of Supernovae

\item Hydrogen-Poor Core-Collapse Supernovae

\item Nebular spectra of supernovae

\item Explosion Physics of Core-Collapse Supernovae

\item Neutrino Driven Explosions

\item Non-spherical initial stellar structure and core collapse

\item Dynamical Evolution and Radiative Processes of Supernova
  Remnants

\item Galactic and Extragalactic Samples of Supernova Remnants: How They Are
Identified and What They Tell Us

\item Supernova remnant from SN1987A

\end{itemize}

\printindex

\end{document}